# Proximity Ascertainment Bias in Early Covid Case Locations

M. B. Weissman

*Department of Physics, University of Illinois at Urbana-Champaign 1110 West Green Street, Urbana, IL 61801-3080, USA*

*mbw@illinois.edu*

Abstract: A comparison of the distances to the Huanan Seafood Market of early Covid cases with known links to the market versus cases without known links shows results apparently incompatible with a location model lacking proximity ascertainment bias. The sign of the difference instead agrees with a model in which such ascertainment bias is large. In the presence of such bias inferences based on the clustering of case locations become unreliable.

Understanding how the SARS CoV-2 virus arose would shed light on whether current pathogen research practices ought to be changed. (*1*) One of the basic starting questions is whether the virus first spilled over into humans from wildlife trade at the Huanan Seafood Market (HSM). Perhaps the strongest piece of evidence supporting the HSM account is presented in a paper by Worobey et al.(*2*) describing geographical clustering of early cases near the HSM. The significance of that clustering has been much disputed based on fundamental issues in spatial techniques(*3*) as well as on a collection of anecdotal reports indicating strong proximity bias in case ascertainment.(*4*) At various stages in the early pandemic contact with wet markets and contact with other known cases were part of the official diagnostic criteria(*5, 6*), with both criteria raising the possibility that detected cases would cluster nearer to a market than the overall population of cases. It is unclear how much these criteria were used unofficially at other stages.(*6*)

It seems unlikely that competing anecdotal stories about how the early cases were found will be persuasive to anyone who already has formed an opinion either way on whether the case locations are important evidence. In what follows I argue that Worobey et al. (*2*) includes *internal* evidence that strongly indicates that there was major proximity ascertainment bias.

Consider two hypotheses, *W* and *M*. *W* is that all the cases ultimately come from the HSM and that accounts for the observed clustering near the HSM, with no major ascertainment bias. *M* is that the proximity ascertainment bias is too large to allow inference about the original source from the location data. These hypotheses have opposite implications for the correlation between detected linkage to HSM and distance from HSM.

For hypothesis *W*, there is some typical distance from HSM to a linked case. An unlinked case must come from a linkable case (typically not observed) via some additional transmission steps in which the traceability is lost. The mean-square displacement (MSD) from the HSM of the unlinked cases would then be approximately the sum of the MSD of the linked cases and the MSD from the linked cases of the remaining steps in which traceability was lost. Barring some unknown contrivance, the unlinked cases would then on average be farther away from HSM than the linked cases. The linkage-distance correlation would be negative, i.e. the probability of a case being linked would decrease with distance from the HSM.

For hypothesis *M* either linkage or proximity can lead to case observation. Some cases would be found by following links, some others by scrutiny near HSM. Unlinked cases would be less likely to be found unless they were near to HSM. Observation would thus be a causal collider between linkage and proximity: linkage→observation←proximity. Then within the observed stratum of cases collider stratification bias(*7*) would give a negative correlation between linkage and proximity, i.e. a positive linkage-distance correlation.

The relevant observational results are reported clearly by Worobey et. al. (*2*): "(ii) cases linked directly to the Huanan market (median distance 5.74 km…), and (iii) cases with no evidence of a direct link to the Huanan market (median distance 4.00 km)…. The cases with no known link to the market on average resided closer to the market than the cases with links to the market ($P = 0.029$)." The statistical significance of the deviation for the *W* hypothesis is even stronger than the "p=0.029" suggests since that p-value is for the hypothesis of no difference but *W* implies a noticeable difference of the opposite sign to the observed effect. The *W* hypothesis is therefore strongly disconfirmed by the data of Worobey et al.(*2*) The sign of the correlation instead agrees with the M hypothesis, that there is substantial proximity-based detection bias.

If hypothesis *W* holds, one would also expect the more distant unlinked cases to be displaced in roughly the same directions as the linked cases from which they descend. Visual inspection of the map of linked and unlinked cases, Fig. 1A(*2*), does not support such an interpretation since the linked cases tend to be north of HSM and the unlinked south and east. Quantitatively, using the case location data from the Supplement(*2*) one finds an angle of 105° between the displacements from HSM to the centroids of the linked and unlinked cases, i.e. a slightly negative dot product between those typical displacements. That would be surprising for any account in which cases start with a linkable market transmission and then at some point lose linkage through an untraceable transmission.

Worobey et al.(*2*) do propose a distinction between types of linked cases that could in principle lead to a difference in distances between linked and unlinked cases. They point out that cases linked because the patient (or a linkable contact) worked at HSM might typically be farther than ones linked via shopping at HSM. They do not directly explain how that would give a distance contrast between linked and unlinked cases. In order for that difference within the *linked* cases to show up as the observed contrast with the u*nlinked* cases one would need to add another assumption, that it would be harder to trace secondary connections to the nearby shoppers. No explanation is given for why such an effect would be expected or for what sign it would be expected to have. One might expect that among people near the market it would be easier to find contacts with neighbors with linked cases. An account of this sort would also need to explain why the more distant unlinked cases are not displaced from HSM in the same general direction as the more distant linked cases. A second-order explanation along these lines remains in the realm of possibility although more speculative than the simple first-order observational collider bias.

The probable existence of major proximity detection bias should not be taken to imply that there is no actual clustering of the unlinked cases. It does mean that these data provide no reliable way of knowing if there is.

Funding: No funding was used for this work.

Conflict of Interest: The author has no conflict of interest.

Data Availability: The spread sheet with the data and calculations for the angle between the centroid displacements is available at

https://drive.google.com/file/d/1sd1kBPRjVSgvKg0gcWoW0MW9tE-9kc1j/view?usp=sharing

A supplemental talk may be downloaded from https://docs.google.com/presentation/d/1y3_KYMnU7YyMjkMvIWcpOxGZkWTfQ6b2/edit?usp=drive_link&ouid=106543307891005183746&rtpof=true&sd=true.

 It can be played in PowerPoint, not Google Docs.

**Addendum (not in the published JRSSA article):**

**Response to a Reply from Débarre and Worobey.**

Débarre and Worobey have posted a Reply(8) to my JRSSA paper(9) on the arXiv. Essentially, they now dismiss the p-value of 0.029 for the linked/unlinked displacement contrast given in the original Worobey et al. paper(*2*). They say that the original p-value may underestimate the chance of finding unlinked cases closer to HSM than linked cases when one takes into account the overdispersion of the SARS-CoV-2 transmission. They present some simulations of multi-step transmission to support the claim that the p-value of the reported results is not as low as initially estimated. They emphasize the possibility I had raised that the typical distance between the point of acquiring an infection and passing it on could be less than the typical distance from the point of acquisition to the point of residence. They claim that their simulation is conservative because its first step goes from HSM to a residence, before subsequent shorter infection-to-infection steps. Since the step operations commute, however, this is exactly equivalent to a more realistic model of infection-to-infection steps followed by a single required step to the reported residence location. (The simulation code used by Débarre and Worobey is peculiar in some respects, e.g. using a probability density function for vector displacement in 2-D that diverges as the displacement approaches zero, but here I will not explore such details.)

The p-values they present are for the unlinked median distance simply being less than the linked rather than for results as extreme as the actual result (a ratio of 0.70) reported by Worobey et al. (*2*) Jon Weissman has re-run the simulation code provided by Débarre to see what p-values it gives for the actual reported data as a function of the ratio of infection-infection distance to infection-residence distances. I.e. he ran the Débarre code with no changes at all other than using the observed 0.70 distance ratio(*2*) rather than an arbitrary 1.0 ratio to calculate the p-value of the observation. The results are in the figure below.

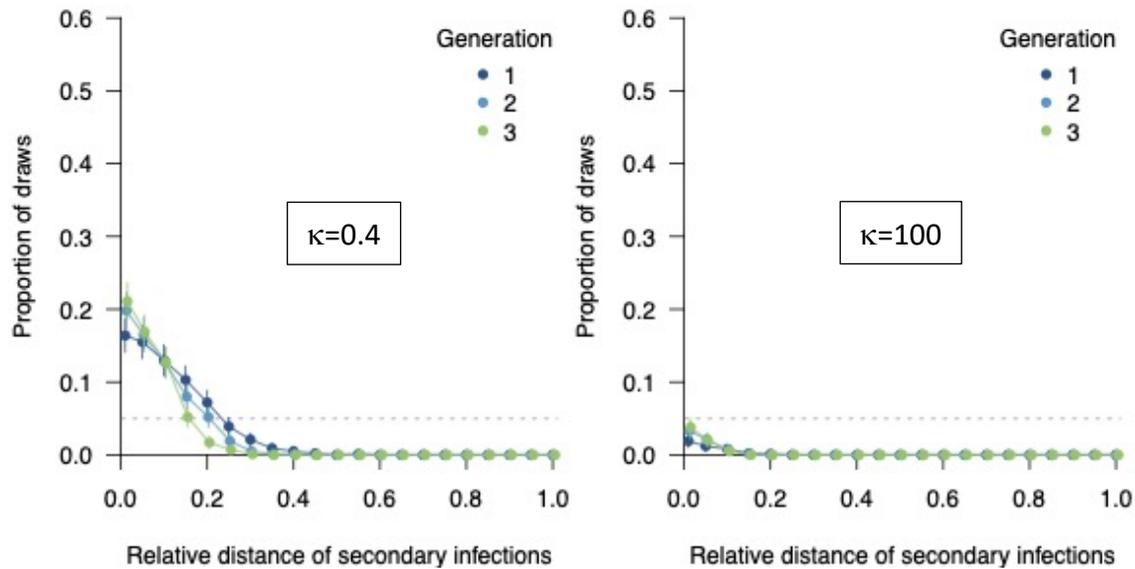

The formatting and labelling are produced directly by the Débarre code except for the labels of the dispersion parameter κ. The x-axis parameter is the ratio of a typical infection-to-infection distance to an infection-to-home distance. The y-axis is the p-value. The generation number gives the number of steps before the final step to the reported residence. Given the ratio of the number of ascertained unlinked cases to ascertained linked cases, the generation number would have to be at least about 2, so the generation 2 and 3 results are the most relevant.

For the nearly Poisson case (κ=100) with no infection-infection displacements (i.e. the y-axis intercept) the p-value approximately reproduces that of the original paper. Within that κ=100 model, apparently used for the original p-value calculation, the probability of the observed ratio is under 0.04 for any parameter value and negligible for any realistic value.

For the more realistic over-dispersed case (κ =0.4) the reported results remain highly unlikely to be consistent with the model unless the typical distance from the place of acquiring the infection to the place of passing it on is less than about 20% of the distance to home. Débarre and Worobey (8) mention "shops, restaurants, bars, and other locations" as places outside the market where infections could be transmitted. An obvious additional category would be public transportation to and from work. Given how much of daily life is spent either at work (i.e. in the market, where spread would be linked rather than unlinked) or near home or in transit, it seems quite unlikely that the displacements to non-work transmission sites would be so small compared to the commute distances. The probability of such a constraint cannot, of course, be derived as a p-value from any formal model.

Débarre and Worobey (8) describe another perhaps more relevant possibility, that the typical distance from infection site to home might be quite a bit smaller for secondary infections than for those starting in the market. If that possibility is combined with a large fraction of the transmission from linkable cases to the unlinkable ones occurring near (but not in) the market rather than near home or on the commute and with the next round or two of transmission also having relatively small displacements, then the observed 0.70 ratio might be explained. Although one clinic-derived infection fitting the pattern is mentioned no evidence is given for that chain of constraints on parameters being common. It remains possible but less parsimonious than simple collider bias.

Thus the linked/unlinked distance contrast can be made reasonably consistent with the $W$ hypothesis by adding in some hypothetical population structure, especially major differences in travel patterns of different groups with different relations to the market spread coupled with enhanced stochasticity due to overdispersion. As Stoyan and Chiu(3) pointed out, omission of such complications was at the heart of the Worobey et al. (2) arguments rejecting a non-market origin by rejecting a highly simplified null model of prevalence tracking population density. Without such real-world complications the reported data look inconsistent with the $W$ hypothesis but with real-world complications the arguments against alternative sources unravel.

Débarre and Worobey (8) also point out that Worobey et al. (2) included a discussion of ascertainment bias in their supplementary material. Most of that discussion is irrelevant to the type of bias that I discuss. The bias specific to the unlinked cases is discussed only in the following passage:

> "So, could those unlinked cases have been detected via biased case-finding involving searching for cases only in neighborhoods near the market but not in other parts of Wuhan? Not likely. Remembering that all the cases were hospitalized and that no diagnostic test was available to identify mild cases it is likely that most or all Huanan market-unlinked cases were ascertained while in hospitals." (*2*)

The argument only applies to the possibility of false positives. The issue, however, is not false positives but something more like false negatives. Unlinked cases away from the market appear likely to have been detected and reported at a disproportionately low rate, just as Bahry (*6*) and others(*4*) had argued based on statements from Wuhan participants.

The possibility of ascertainment bias is not a post-hoc hypothesis to explain the negative correlation between linkage and distance from HSM. It was independently proposed by many of those most aware of the data collection issues. For example, virologist Jeremy Farrar writes "That tight case definition resulted in an Escher's loop of misguided circular reasoning: testing only those people with a link to the market created the illusion that the market was the source of the disease…" (10)